\documentclass[a4paper,11pt]{article}

\usepackage{jcappub} 

\usepackage[T1]{fontenc} 
\usepackage{amsmath}
\usepackage{siunitx}
\usepackage{hyperref}

\DeclareSIUnit \parsec {pc}

\title{\boldmath Analytical perturbations of relativistic images in Kerr space-time}

\author[a,b]{Fabio Aratore}
\author[a,b]{and Valerio Bozza}

\affiliation[a]{Dipartimento di Fisica “E.R. Caianiello”, Università degli studi di Salerno, Via Giovanni Paolo II 132, I-84084 Fisciano SA, Italy}
\affiliation[b]{Istituto Nazionale di Fisica Nucleare, Sezione di Napoli,
Via Cintia, 80126 Napoli NA, Italy}

\emailAdd{faratore@unisa.it}
\emailAdd{valboz@sa.infn.it}

\abstract{Light rays passing very close to black holes may wind several times before escaping. For any given electromagnetic source around the black hole, a distant observer would thus observe two infinite sequences of images on either side of the black hole. These images are generated by light rays performing an increasing numbers of loops. The strong deflection limit provides a simple analytic formalism to describe such higher order images for spherically symmetric metrics, while for axially symmetric black holes one typically resorts to numerical approaches. Here we present the leading order perturbation to higher order images when the black hole spin is turned on. We show that the images slide around the black hole shadow as an effect of space-time dragging. We derive analytical formulae for their shifts and the perturbation of their time delays. We also discuss how such simple analytical formulae for images by Kerr black holes can be of great help in many applications.}

\notoc

\begin{document}
\maketitle
\flushbottom

\section{Introduction}
Gravitational lensing refers to the collection of effects, consequences of Einstein's General Relativity, that occur when the gravitational field of a massive object bends and distorts the trajectories of light rays passing nearby. Particularly interesting is the case of black holes and some other compact objects that can cause arbitrarily large deflections \cite{darwin1959gravity, Ohanian1987, Atkinson-1965, falcke1999viewing, falcke2000shadow}. 

As a consequence, these deflections lead to the creation of two infinite sequences of images of the same point source on both sides of the black hole with each successive image exponentially dimmer and closer to the border of a dark region called shadow \cite{perlick2022calculating}. The additional images, called relativistic or higher-order images, have been extensively studied from both numerical and analytical perspective \cite{virbhadra2000schwarzschild, bozza2001g, eiroa2002reissner, bozza2002gravitational, BozzaMancini2004, eiroa2004strong, bozza2005analytic, bozza2006kerr, bozza2007strong,bozza2010}.

Initially, this phenomenon was considered a mere theoretical curiosity for decades because there seemed to be no hope of observing such faint signals. However, with the advent of modern observational technologies, gravitational lensing of compact objects has evolved into an invaluable tool for unraveling the mysteries of our cosmos. In fact, since the Event Horizon Telescope (EHT) has reached the sensitivity for supermassive black holes horizon-scale measurements and revealed the existence of the shadow surrounded by an emission ring due to the luminous matter of the accretion disk \cite{akiyama2019first1, akiyama2022first1}, the studies about relativistic images, in this case called photon rings since they replicate the whole accretion flow in very thin rings, have gained a great practical importance in studying the astrophysics of black hole environment. \cite{gralla2019black, gralla2020lensing, gralla2020observable, Gan-Wang-2021, wielgus2021photon, Hadar2021, ayzenberg2022testing, broderick2022photon, papoutsis2023jets, broderick2023shadow}.

While higher-order images remain beyond the observational capabilities of current interferometry with terrestrial baselines, some projects aiming for the extension of very long baseline interferometry technique (VLBI) seem feasible within the next decade \cite{pesce2019extremely, pesce2021toward, andrianov2021simulations, Andrianov-2022}. They predict that (sub)millimeter interferometric observations with $\simeq \SI{0.1}{\mu as}$ resolution and $\simeq \SI{1}{\mu Jy}$ sensitivity could access more than a million  supermassive black hole shadows and even the first and second-order photon rings. 

These rings seem to contribute with negligible signal but Refs.~\cite{johnson2020universal} and \cite{gralla2020shape} show that these sub-rings produce strong and universal signatures in the complex visibility function obtained by long interferometric baselines. Such features, being relatively insensitive to the astrophysical details of the emitting distribution of matter, present a unique opportunity for precise measurements of black hole parameters, such as mass and spin, in addition to offering a means of testing the principles of General Relativity. Remarkably, such measurements can be achieved using only a sparse interferometric array, representing a promising avenue for advancing our understanding of black holes and fundamental physics \cite{Himwich2020, Broderick-2022-spin, eichhorn2023universal, staelens2023black, kocherlakota2023extreme, da2023photon}.

Analytical calculations of higher-order images are possible through the use of the strong deflection limit technique (SDL), which has been proposed many times \cite{darwin1959gravity,Atkinson-1965,Luminet1979,Ohanian1987} and lately revisited and extended to any spherically symmetric metrics \cite{bozza2002gravitational,bozza2007strong}. This method provides a simple logarithmic expression for large deflection angles particularly suited for photons that complete at least one full orbit around the black hole \cite{bozza2010}.

The SDL has been applied and investigated to a multitude of space-time metrics derived in different gravitational theories (see e.g. \cite{Bhadra2003,Whisker2005,Eiroa2005,Eiroa2006,Ovgun2018,Tsukamoto2016,Tsukamoto2017,Chen2009,Bin-Nun2010,Tsukamoto2021,Iyer2007,Chakraborty2017JCAP...07..045C}). With the use of SDL, Refs.~\cite{bisnovatyi2008strong,tsupko2013gravitational,bisnovatyi2022analytical,tsupko2022shape} analytically investigate the properties of higher-order images originating from an accretion disk or a thin emission ring surrounding a Schwarzschild black hole. In addition, Ref.~\cite{Aldi-Bozza-2017} studies the shapes of relativistic iron lines observed in spectra of black holes. In Ref.~\cite{aratore2021decoding}, we studied the possibility of analytically extracting information about the metric from the interferometric patterns due to relativistic images of a compact source in a generic spherically symmetric space-time.

However, spherically symmetric black holes represent an idealized limit, since all astrophysical objects spin. Simple considerations on the conservation of angular momentum lead to the conclusion that all astrophysical black holes are expected to have a non-negligible angular momentum. Extensions of the SDL to rotating black holes have been attempted in several papers \cite{Vazquez2004,bozza2005analytic,bozza2006kerr,bozza2007strong,Chen2012,Wei2012,Islam2021,Barlow2017,tien2021strong,tien2021time}, focusing on some specific limits.  

In this article, our focus is directed towards the perturbations of relativistic images originating from a point light source within the framework of Kerr space-time in the limit of small angular momentum. We present entirely analytical formulas that describe the deviations in the positions and time delays of these images in comparison to the spherically symmetric scenario. Moreover, we establish a direct connection between these perturbations and the features of a rotating black hole. To our knowledge, in spite of decades of investigations of geodesics in Kerr space-time, such analytical perturbative formulae have not been derived yet and demonstrate the power of the SDL technique once more.

The structure of this paper is organized as follows: In Sec.~\ref{Kerr null geodesics} and \ref{The shadow of a black hole}, we provide a summary of well-known properties of null geodesics in Kerr black holes, including the form of the photon sphere and the shadow as function of the polar angle in the observer sky. Sec.~\ref{The strong deflection limit} contains the basis of SDL technique.  Subsequently, we investigate the position and time delay of relativistic images, initially starting from the spherically symmetric case and then incorporating perturbations induced by rotation to the leading order in the specific angular momentum. Finally, we draw our conclusions, summarizing the key findings and implications of our study.

\section{Kerr null geodesics}
\label{Kerr null geodesics}
Let us consider the Kerr metric, which in Boyer-Lindquist coordinates $x^\mu\equiv \left(t, r, \theta,\phi\right)$ reads
\begin{equation}
\begin{split}
    ds^2=\frac{\Delta-a^2\sin^2{\theta}}{\rho^2}&c^2 dt^2-\frac{\rho^2}{\Delta}dr^2-\rho^2 d\theta^2+ \\ -\frac{(r^2+a^2)^2-a^2\Delta\sin^2{\theta}}{\rho^2}&\sin^2{\theta}d\phi^2+\frac{4mar\sin^2{\theta}}{\rho^2}dt d\phi
\end{split}
\end{equation}
where
\begin{gather}
    \Delta=r^2-2mr+a^2 \\
    \rho^2=r^2+a^2\cos^2{\theta}.
\end{gather}
Here, the parameter $m=GM/c^2$ represents the conventional normalized mass parameter, where $M$ stands for the mass of the black hole. Additionally, $a=L/Mc$ designates the specific angular momentum, which ranges from 0 (corresponding to the Schwarzschild black hole) to $m$ for the extremal Kerr black hole.

\begin{figure}
\centering
\includegraphics[width=0.7\linewidth]{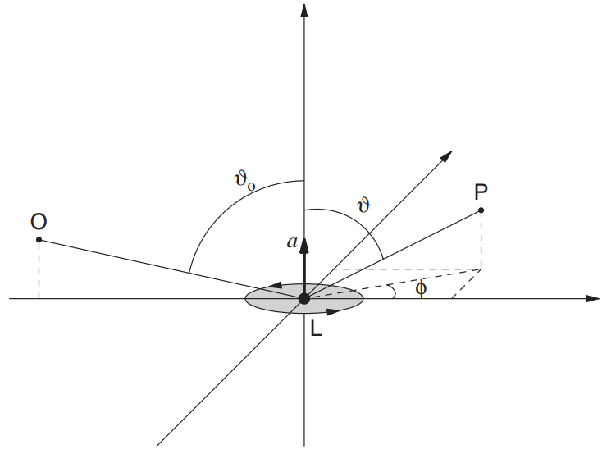} 
\caption{Boyer-Lindquist coordinates in Kerr metric (also
called spin-oriented coordinates). $L$ represent the black hole with its spin parameter $a$, $O$ the observer and $P$ a generic point in this space-time. The gray disk visualizes the equatorial plane.}
\label{BLcoordinates}
\end{figure}

We consider a static point-source located at coordinates $(r_s, \theta_s, \phi_s)$, whereas the observer is at $(r_o, \theta_o, \phi_o)$. For the sake of convenience, we opt to set the azimutal angle $\phi$ such that $\phi_o = \pi$, as illustrated in Figure~\ref{BLcoordinates}. Moreover, throughout the text we will use the notation $\mu=\cos{\theta}$.

Carter demonstrated that geodesics can be integrated by explicit quadratures \cite{carter1968global}. In particular, the null geodesics can be expressed as
\begin{equation}
     \int \frac{d r}{\sqrt{R}}=\int \frac{d\theta}{\sqrt{\Theta}}
     \label{geo1}
\end{equation}
\begin{equation}
        \phi_f -\phi_i=a\int \frac{r^2+a^2-a J}{\Delta\sqrt{R}}d r-a\int \frac{d r}{\sqrt{R}}+J\int\frac{\csc^2{\theta}}{\sqrt{\Theta}}d\theta
        \label{geo2}
\end{equation}
\begin{equation}
        ct=\int \frac{r^2}{\sqrt{R}}d r+a^2\int \frac{\cos^2{\theta}}{\sqrt{\Theta}}d\theta +2m\int\frac{r(r^2+a^2-a J)}{\Delta\sqrt{R}}d r
        \label{geo3}
\end{equation} 
where
\begin{equation}
    \Theta=Q+a^2 \cos^2{\theta}-J^2\cot^2{\theta}
\end{equation}
\begin{equation}
    R=r^4+\left(a^2-J^2-Q\right)r^2+2mr\left[Q+\left(J-a\right)^2\right]-a^2 Q.
    \label{R}
\end{equation}
We indicate the initial and the final value of the azimutal coordinate of the photon with $\phi_i$ and $\phi_f$ respectively. Obviously, in our treatment we will consider $\phi_i$ as the azimut of the source and $\phi_f$ as the azimut of the observer.

Null geodesics in Kerr space-time coming from an infinite distance can be distinguished in two classes: geodesics crossing the event horizon and ending in the central singularity; geodesics that do not cross the horizon but reach a closest approach distance and then go back to infinity where they can eventually reach a distant observer.

In a spherically symmetric space-time, there exists a specific distance denoted with $r_m$, which correspond to an unstable circular orbit. This means that $r_m$ also represents the minimum of the closest approach distance for which incoming photons can still escape from the black hole. If they move any closer to the black hole, their path is inevitably doomed to terminate into the central singularity. For this reason $r_m$ is commonly referred to as the radius of the photon sphere. In the contest of Kerr metric, because of the effects of rotation, it is expected that prograde light rays are allowed to get closer to the black hole. As a result, there is no unique $r_m$ but rather a family of radii depending on the direction of arrival of the photon.

In practice, the distinction between the above mentioned paths is done considering whether the constants of motion $J$ and $Q$ lie inside or outside the critical locus $(J_m,Q_m)$. The equations determining the critical locus are $R(r)=0$ and $R'(r)=0$. We then obtain the critical locus  parameterized by $r_m$:
\begin{equation}
    J_m=\frac{r_m^3-3m r_m^2+a^2r_m+m a^2}{a(m-r_m)}
    \label{criticallocus1rm}
\end{equation}
\begin{equation}
    Q_m=-\frac{r_m^3(r_m^3-6m r_m^2+9m^2 r_m-4m a^2)}{a^2(m-r_m)^2}.
    \label{criticallocus2rm}
\end{equation}

The minimum and maximum value of $r_m$ are determined by the equation $Q_m=0$, which is a third degree polynomial.

\section{The shadow of a black hole}
\label{The shadow of a black hole}
The position in the sky where the observer detects the photon is closely related to the constants of motion $J$ and $Q$. In particular for an observer in the asymptotic flat region ($r_o\gg m$) this relation is particularly simple. In fact, choosing an orthogonal angular coordinate system $(\theta_1,\theta_2)$ on the observer's sky centered on the black hole, with the axis $\theta_2$, aligned with the projection of the black hole's spin onto the sky, we can express this relation as follows \cite{perlick2022calculating,chandrasekhar1998mathematical}:
\begin{equation}
    r_o\theta_1=-\frac{J}{\sqrt{1-\mu_o^2}}
    \label{observersky1}
\end{equation}
\begin{equation}
    r_o\theta_2=\pm\sqrt{Q+\mu_o^2\left(a^2-\frac{J^2}{1-\mu_o^2}\right)}.
    \label{observersky2}
\end{equation}
These relations can be used to convert the critical locus in the space of the constants of motion, into a new one $(\theta_{1 m},\theta_{2 m})$ in the observer sky, also parameterized with $r_m$. This particular locus has been commonly called the shadow of the black hole \cite{perlick2022calculating} because, as previously mentioned, photons within the region $(J_m,Q_m)$ are destined to cross the event horizon. In our case, this means that geodesics appearing in the observer's sky within the shadow boundary have been emitted by some foreground sources or by matter falling in the black hole just before it crosses the horizon. In the case of electromagnetic emission from an accretion disk, the main dark area extends to the lensed position of the inner edge of the disk \cite{Luminet1979,beckwith2005extreme}. Therefore, if the emission extends all the way down to the event horizon, the dark area is much smaller than the shadow boundary as defined above \cite{Chael2021}. Therefore, the locus $(\theta_{1 m},\theta_{2 m})$ has also been dubbed ``critical curve'' in recent papers to distinguish it from the ``inner shadow'', which corresponds to the dark area in the famous EHT images. Yet, this term again introduces some confusion with critical curves of the lens map, which have been traditionally defined as the loci where the Jacobian of the lens map vanishes. In the absence of a more appropriate definitive term, in this paper we will still use the wording ``shadow boundary'' to make contact with the vast majority of previous papers \cite{perlick2022calculating}.

The radius of the unstable circular orbit (photon sphere) is fixed to $r_m=3m$ for Schwarzschild metric. In contrast, for Kerr black holes, $r_m$ can vary between two limiting values, depending on the direction of the incoming photon. As anticipated at the end of the previous section, these two extreme values can be obtained by solving the equation $Q_m=0$, which is a cubic equation in $r_m$. It is now time to introduce our perturbative expansion in $a$. Our purpose is to study the perturbations of higher order images for small angular momenta. Therefore, here we calculate the two limiting values expanded to the third order in $a$ \cite{bozza2006kerr}:
\begin{equation}
    r_{m \pm}=3m\pm\frac{2}{\sqrt{3}}a-\frac{2}{9}\frac{a^2}{m}\pm\frac{5}{27\sqrt{3}}\frac{a^3}{m^2}+o(a^4).
    \label{rmlimit}
\end{equation}

The locus $(J_m,Q_m)$ is formed by trajectories with $r_m$ between these two boundaries, as calculated by Eqs.(\ref{criticallocus1rm})-(\ref{criticallocus2rm}). Correspondingly, each trajectory will correspond to a point in the shadow boundary through Eqs.(\ref{observersky1})-(\ref{observersky2}). It is convenient to convert the parameterization in $r_m$, which is completely obscure to the observer, with a variable that makes immediate contact with what the observer sees on the sky. We choose the zero-order position angle $\alpha$ along the shadow boundary, defined by
\begin{equation}
\tan{\alpha}=\frac{\theta_{2m}}{\theta_{1m}}.
\end{equation}
Thus, starting from the value of the photon sphere radius in Schwarzschild metric $3m$ and adding perturbations up to the third order in $a$ we obtain the following form:
\begin{equation}
    \begin{split}
        r_m(\alpha)=3m+\frac{2}{\sqrt{3}}a\sqrt{1-\mu_o}\cos{\alpha}-\frac{2}{9}\frac{a^2}{m}\left[2-\left(1-\mu_o^2\right)\cos^2{\alpha}\right]
    \\ +\frac{1}{27\sqrt{3}}\frac{a^3}{m^2}\sqrt{1-\mu_o}\cos{\alpha}\left[10+\mu_o^2-5\left(1-\mu_o^2\right)\cos^2{\alpha}\right]+o\left(a^3\right).
    \label{rmalfa}
    \end{split}
\end{equation}
As the parameter $\alpha$ varies in the interval $[0,2\pi]$, this formula gives all possible values of $r_m$. In particular, for equatorial observer ($\mu_o=0)$, setting $\alpha=0, \pi$ we recover the two limiting values given in Eq.\eqref{rmlimit}. In contrast, for a polar observer, namely $\mu_o=1$, the radius of the photon sphere is slightly smaller but, since we are looking at the black hole from the axis of symmetry and thus all directions are equivalent, $r_m$ does not depend on $\alpha$ anymore:
\begin{equation}
    r_m(\alpha)=3m-\frac{4}{9}\frac{a^2}{m}.
\end{equation}

Moreover, this parameterization allows us to rewrite the critical values of the constants of motion $J_m$ and $Q_m$ as a series expansion in the specific angular momentum:
\begin{equation}
    \begin{split}
        J_m(\alpha)=-3\sqrt{3}m\sqrt{1-\mu_o^2}\cos{\alpha}-2a(1-\mu_o^2)\cos^2{\alpha} \\ +\frac{1}{6\sqrt{3}}\frac{a^2}{m}\sqrt{1-\mu_o^2}\cos{\alpha}\left[3+(1-\mu_o^2)\sin^2{\alpha}\right]+o(a^3)
        \label{criticallocus1}
    \end{split}
\end{equation}
\begin{equation}
    \begin{split}
            Q_m(\alpha)=\left(27m^2+12\sqrt{3}m a\sqrt{1-\mu_o^2}\cos{\alpha}\right)\left[1-(1-\mu_o^2)\cos^2{\alpha}\right] \\ -a^2\left[4-\left(1-\mu_o^2\right)\cos^2{\alpha}\left(9-\mu_o^2-5(1-\mu_o^2)\cos^2{\alpha}\right)\right]+o(a^3)\quad
             \label{criticallocus2}
    \end{split}
\end{equation}
and, recalling the relations Eq.\eqref{observersky1} and Eq.\eqref{observersky2}, we can express the shadow border parameterized in $\alpha$ as follows: 
\begin{equation*}
    \theta_{1m}(\alpha)=\theta_m(\alpha)\cos{\alpha} \qquad \theta_{2m}(\alpha)=\theta_m(\alpha)\sin{\alpha} \label{shadowborder}
\end{equation*}
with
\begin{equation*}
    r_o \theta_m(\alpha)=3\sqrt{3}m+2a\sqrt{1-\mu_o^2}\cos{\alpha}-\frac{1}{6\sqrt{3}}\frac{a^2}{m}\left[4-\mu_o^2-\left(1-\mu_o^2\right)\cos^2{\alpha}\right]
\end{equation*}

It is evident that in the Schwarzschild case ($a=0$), the previous relations correspond to the parameterization of a circle centered in $(0,0)$ and with radius equal to $3\sqrt{3}m$. As the black hole's spin increases, the shadow undergoes a slight distortion and shifts to the right. This implies, as expected, that prograde photons, approaching from the left side as seen from the observer, are allowed to get closer to the black hole, while retrograde photons must keep farther.

When the observer is positioned along the axis of symmetry (i.e. $\mu_o=1)$, the shadow is once more centered in $(0,0)$ and the rotation slightly decreases its radius:
\begin{equation}
    \begin{split}
        r_o\theta_{1m}(\alpha)=\left(3\sqrt{3}m-\frac{1}{2\sqrt{3}}\frac{a^2}{m}\right)\cos{\alpha}+ o\left(a^3\right)
    \end{split}
\end{equation}
\begin{equation}
    \begin{split}
        r_o\theta_{2m}(\alpha)=\left(3\sqrt{3}m-\frac{1}{2\sqrt{3}}\frac{a^2}{m}\right)\sin{\alpha}+ o\left(a^3\right).        
    \end{split}
\end{equation}

With respect to the parameterization with the variable $\xi$ used in Ref.~\cite{bozza2006kerr}, the parameterization in terms of the position angle $\alpha$ is much more advantageous for the study of the perturbations of the images as it is directly connected with the coordinates of the images on the sky.

\section{The strong deflection limit}
\label{The strong deflection limit}
In the investigation of gravitational lensing within the framework of the strong deflection limit (SDL), our focus is on light rays with closest approach lying in a narrow shell situated immediately outside the photon sphere. These light rays experience large deflections.

Specifically, the SDL approximation is particularly well-suited for deflection angles exceeding $2\pi$. In other words, it is ideal for light rays that complete at least one full orbit around the black hole. The images produced by such photons are referred to as relativistic images, and these are the only ones we consider \cite{bozza2010}. 

It is useful to introduce the following parameterization: 
\begin{equation}
    \begin{cases}
        \theta_1(\alpha,\epsilon)= \theta_{m}(\alpha) \left(1+\epsilon\right) \cos \alpha \\
        \theta_2(\alpha,\epsilon)= \theta_{m}(\alpha) \left(1+\epsilon\right) \sin \alpha.
    \end{cases}
    \label{sdlparametrization0}
\end{equation}
Varying $\alpha$ in the range $[0,2\pi]$ and $\epsilon$ in the range $[-1,+\infty]$, the previous parameterization cover the whole observer sky, because $\alpha$ is the polar angle measured counterclockwise with respect to the positive direction of $\theta_1$-axes, while $\epsilon$ fixes the angular distance from the center of the black hole. 

As already specified, in the spirit of SDL, the parameter $\epsilon$ is considered to be small and positive so that, in performing the integrals of the geodesics, we will keep only terms to the lowest order in $\epsilon$.

With the parameterization Eq.\eqref{sdlparametrization0} and the relations Eqs.\eqref{observersky1} and \eqref{observersky2}, it is easy to find the expression of the constants of motion $J$ and $Q$ for these strongly deflected photons:
\begin{equation}
    \begin{cases}
        J(\alpha,\epsilon)= J_m(\alpha) \left(1+\epsilon\right) \\
        Q(\alpha,\epsilon)= Q_m(\alpha)+2\epsilon \left(Q_m(\alpha)+a^2 \mu_o^2\right) +o(\epsilon^2).
        \label{sdlparametrization1}
    \end{cases}
\end{equation}
Substituting these expressions containing $\epsilon$ in the function $R$ given in Eq.\eqref{R} and solving the equation $R=0$ for the closest approach distance we get
\begin{equation}
        r_0(\alpha,\delta)=r_m(\alpha)(1+\delta)
        \label{sdlparametrization2}
\end{equation}
where
\begin{equation}
    \small
            \delta=\sqrt{\frac{2\epsilon}{3}}\left[1+\frac{1}{3\sqrt{3}}\frac{a}{m}\sqrt{1-\mu_o^2}\cos{\alpha}+\frac{1}{54}\frac{a^2}{m^2}\left(8+\mu_o^2-12(1-\mu_o^2)\cos^2{\alpha}\right)\right].
        \label{sdlparametrization3}
\end{equation}
As $\epsilon$ represents the separation of the direction of the incoming photon from the shadow of the black hole, $\delta$ represents the relative difference between the closest approach distance $r_0$ and its minimum possible value $r_m$. It will be synthetically called ``the approach parameter''.

We might expect that in the limit of large positive values of $\delta$, photons would exhibit weak deflection, leading to the classical weak gravitational lensing theory. Conversely, approaching the radius of the photon sphere, for $\delta\to 0^+$, the deflection angle diverges. It's worth noticing that the relation between $\delta$ and $\epsilon$ guarantees that the SDL can be equivalently formulated in terms of either of these two parameters.

In fact, inverting the previous relation, we have:
\begin{equation}
    \epsilon=\frac{3}{2}\delta^2\left[1-\frac{2}{3\sqrt{3}}\frac{a}{m}\sqrt{1-\mu_o^2}\cos{\alpha}-\frac{1}{27}\frac{a^2}{m^2}\left(8+\mu_o^2-15(1-\mu_o^2)\cos^2{\alpha}\right)\right] \label{deltatoeps}
\end{equation}
This brief introduction provides all the tools needed to solve the integrals appearing in the geodesic equations in which there are four radial and three angular integrals: the former ones can be solved using the strong deflection limit technique to which we dedicate the appendix \ref{Resolution of radial integrals}, while the latter ones can be solved with a quite straightforward calculation but paying attention to what kind of angular motion is performed by photons. This is extensively explained in appendix \ref{Resolution of angular integrals}.

\section{Zero order geodesics}
\label{sectionzeroordergeo}
\subsection{Deflection}
A good starting point to understand the basic properties of the problem is the Schwarzschild case. Let us begin by considering $a=0$. In order to write more compact expressions, it is useful to define the angle $\alpha_s$ as
\begin{equation}
    \cos \alpha_s = \frac{\sqrt{1-\mu_o^2}}{\sqrt{1-\mu_s^2}} \cos \alpha.
\end{equation}
Whereas $\alpha$ is the position angle of the image in the observer sky, $\alpha_s$ has the same geometric meaning from the source side. Given the symmetry of the problem, it is a useful quantity for the simplification of many expressions.

Furthermore, as discussed in the appendix, we also introduce the notation $\eta=1-r_0/r\simeq 1- r_m/r$ to express the radial distance from the black hole. Subscripts $s$ and $o$ will still identify values assumed for source and observer respectively.

Finally, many expressions in the radial integrals can be simplified by introducing the short notation
\begin{equation}
    S_c(\eta)=\frac{c-\sqrt{3-2\eta}}{c+\sqrt{3-2\eta}}. \label{Sdefinition}
\end{equation}

Collecting the results obtained in the appendices we can express the null geodesics equations \eqref{geo1} and \eqref{geo2} in the following form:
\begin{equation}
    \begin{cases}
        \psi=n\pi\mp\mathrm{arccot}{\frac{\sqrt{1-\mu_s^2} \sin \alpha_s}{\mu_s}}\pm(-1)^n\mathrm{arccot}{ \frac{\sqrt{1-\mu_o^2} \sin \alpha}{\mu_o}}     \\
       \phi_s -\pi = n\pi \mp \arctan \left(\mu_s\cot{\alpha_s}\right) \pm(-1)^n\arctan\left(\mu_o\cot{\alpha}\right) 
    \end{cases}
    \label{zeroordergeo}
\end{equation}
where we have identified the function
\begin{equation}
    \psi= -2\log\delta+ \log\left[12 \ S_{\sqrt{3}}(\eta_s)\right]+\log\left[12 \ S_{\sqrt{3}} (\eta_o)\right].
    \label{psi0}
\end{equation}
This quantity is nothing but the deflection induced by a Schwarzschild black hole measured in the plane where the light ray lies. Following previous works \cite{bozza2006kerr,bozza2007strong}, we will refer to it as the ``scalar deflection". 

We also recall that $n$ indicates the number of inversions in the polar motion while the double signs present in the equations come from the angular integrals and should be treated as follows: If the photon departs from the source by increasing its value of $\mu$, then we should opt for the upper signs, otherwise we will select the lower signs. In addition, from Ref.~\cite{bozza2010} we know that the SDL approximation improves for increasing value of $n$. For this reason we consider only photons that complete at least one loop around the black hole before escaping to infinity, so that the minimum value of $n$ we will consider is 2. 

Once the position of the source $(\eta_s, \mu_s, \phi_s)$ is given, the equations in \eqref{zeroordergeo} can be solved in terms of $\psi$ and $\alpha$. The first variable is connected to $\delta$ and then $\epsilon$ through Eqs.(\ref{psi0})-\ref{deltatoeps}), which gives the fractional radial distance of the image from the shadow, while $\alpha$ has been defined by us as the position angle along the shadow. 

An alternative way to calculate $\alpha$ is by recalling that the entire motion of the photon takes place on a single plane containing the source and the observer. Referring to Fig.~\ref{BLcoordinates}, calling $\hat{O}$ the unitary vector pointing to the observer and $\hat{S}$ the corresponding one for the source, we define a new vector $\hat{K}=\hat{O}\times\hat{S}/\lvert\lvert \hat{O}\times\hat{S}\rvert\rvert$ orthogonal to the geodesics plane defined by $\hat{O}$ and $\hat{S}$; therefore $\Vec{K}$ is also orthogonal to the line of sight. The observer will see the image form in the direction $\hat{I}=\hat{K}\times\hat{O}$. Now, in the $(\theta_1,\theta_2)$ angular coordinate system in the observer sky, the axis of the $\theta_1$ coordinate is orthogonal to the observer direction and spin direction simultaneously. Being $\hat a$ the unit vector along the spin axis, we define the unit vector $\hat \theta_1 = \hat a \times \hat O / \lvert\lvert \hat a \times \hat O\rvert \rvert $. Therefore, the position angle of the image will just be given by $\cos \alpha = \hat I \cdot \hat \theta_1$. 

Through this simple geometric procedure we are able to find the following analytical expression (see also \cite{Kraniotis2011}):
\begin{equation}
    \cos{\alpha}=\mp\frac{\sqrt{1-\mu_s^2}\sin{\phi_s}}{\sqrt{1-\left(\mu_o\mu_s-\sqrt{1-\mu_o^2}\sqrt{1-\mu_s^2}\cos{\phi_s}\right)^2}}.
    \label{alfa0}
\end{equation}
It is easy to check that this expression solves the second equation of (\ref{zeroordergeo}). Once this angle is known, we are able to determine $\psi$ using the first equation in the system \eqref{zeroordergeo}. The double sign appearing in $\cos{\alpha}$ provide the appearance of a couple of images on opposite sides of the black hole shadow, i.e. $\alpha$ and $\alpha+\pi$.

The expression contained in Eq.\eqref{alfa0} does not depend on the index $n$. This means that in spherically symmetric situation all the images, whatever the order, appear at the same position angles on both sides, with a scalar deflection $\psi$ differing from one image to another, implying smaller and smaller distances $\epsilon$ from the shadow border.

\begin{figure}
\centering
\includegraphics[width=0.8\textwidth]{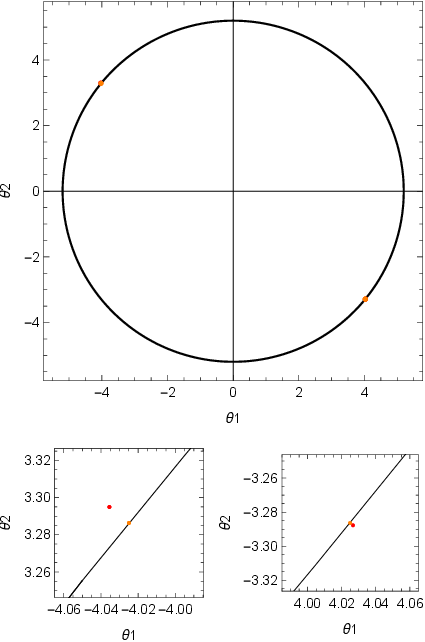}\newline
\caption{An example of relativistic images of a point-like source located at $(\eta_s=0.9, \theta_s=\pi/3, \phi_s=\pi/4)$ with the observer on the equatorial plane. The black circle represent the shadow border, the red and the orange dots are the images characterized by the index $n$ being equal to 3 and 5 respectively.  The lower panels show a zoom on the images showing more clearly the difference between the images created on the same side of the source and the ones created on the opposite side. The angles $\theta_1$ and $\theta_2$ are measured in units of $m/r_o$.}
\label{Schwarzschild}
\end{figure}

In conclusion:
\begin{itemize}
    \item the angle $\alpha$ is recovered from Eq.\eqref{alfa0};
    
    \item  $\psi$ is given by the first equation of \eqref{zeroordergeo}:
\begin{equation}
    \psi=n\pi\mp h(\alpha) \label{psinpi}
\end{equation}
where
\begin{equation}
\small
    h(\alpha)=\mathrm{arccot}{\frac{\sqrt{1-\mu_s^2} \sin \alpha_s}{\mu_s}}-(-1)^n\mathrm{arccot}{ \frac{\sqrt{1-\mu_o^2} \sin \alpha}{\mu_o}};
\end{equation}

    \item the approach parameter $\delta$ can be immediately computed using the definition Eq.\eqref{psi0}:
\begin{equation}
    \delta^2=Ae^{-\psi}=Ae^{-n\pi\pm h(\alpha)}
    \label{eq:deltazeroquadro}
\end{equation}
where
\begin{equation}
    A=144S_{\sqrt{3}} (\eta_s)S_{\sqrt{3}} (\eta_o);
\end{equation}

    \item from Eq.\eqref{sdlparametrization3}, setting $a=0$, we have:
    \begin{equation}
        \epsilon=\frac{3}{2}\delta^2=\frac{3}{2}Ae^{-n\pi\pm h(\alpha)};
    \end{equation}

    \item and finally, through Eqs.\eqref{sdlparametrization0} and \eqref{shadowborder} with $a=0$, the coordinates of the images on the observer's screen are:
    \begin{equation}
         \begin{cases}
         r_o\theta_1=3\sqrt{3}m \cos{\alpha}\left(1+\frac{3}{2}Ae^{-n\pi\pm h(\alpha)}\right) \\
         r_o\theta_2=3\sqrt{3}m \sin{\alpha}\left(1+\frac{3}{2}Ae^{-n\pi\pm h(\alpha)}\right)
    \end{cases}
    \label{eq:posizioneimmaginizero}
    \end{equation}
\end{itemize}
We also use the hypothesis that the observer is infinitely distant ($r_o\gg m)$. Thus, the constant $A$ should be viewed in the limit of $r_o\to\infty$ or $\eta_o\to 1$:
\begin{equation}
    \lim_{\eta_o\to 1} A=144S_{\sqrt{3}} (1)S_{\sqrt{3}} (\eta_s)
\end{equation}
An example of the relativistic images formation can be seen in Fig.~\ref{Schwarzschild}.

\subsection{Time delay}
Let us now  concentrate on the third geodesic equation Eq.\eqref{geo3}. Using again the results from the appendices and setting $a=0$ we obtain:
\begin{equation}
    ct=m\left(3\sqrt{3}\psi-K_s-K_o\right)
    \label{zeroordertimegeo}
\end{equation}
where the constants $K_i$ are:
\begin{equation}
\small
   K_i=2\log\frac{S_{\frac{5}{3}}(0)}{S_{\frac{5-2\eta_i}{3}}(\eta_i)}+3\left(\frac{\sqrt{3-2\eta_i}}{1-\eta_i}-\sqrt{3}\right)
\end{equation}
When integrating the time geodesic equation from the source to the observer's location, this provides the time it takes for a photon to travel between these two points. In fact, the expression in Eqs.\eqref{zeroordergeo} is proportional to $\psi$ since it quantifies the number of completed orbits. In addition, the constants $K_i$ diverges when $\eta_i=1$. However, this is a trivial divergence because $\eta=1$ corresponds to an infinite distance, resulting in an infinitely long travel time. This divergence is not a concern because our interest lies in the time delay between pairs of images, and these constants, which are solely dependent on the positions of the source and the observer, will cancel out in the subtraction.

In this way the time delay between a pair of images, indicating them with the subscript $i$ and $j$, simply becomes:
\begin{equation}
    \Delta t_{i,j}=3\sqrt{3}\frac{m}{c}\left(\psi_i-\psi_j\right).
    \label{delay0}
\end{equation}
Now we have to consider two separate situations: the case of consecutive images appearing on the same side of the black hole and the case of images appearing on different sides of the black hole with the same $n$.

In the first case, the scalar deflection has to be taken with the same sign obtaining the simple expression:
\begin{equation}
   \Delta t_+=6\sqrt{3}\pi\frac{m}{c}.
\end{equation}
This time does not depend on the particular configuration of the system but only on the mass of the black hole. This is essentially because it is the time taken by the photon to cover a full orbit.

In the second case instead, $\psi$ must be taken with different sign, and thus:
\begin{equation}
\footnotesize
    \Delta t_-=6\sqrt{3}h(\alpha)\frac{m}{c}.
\end{equation} 

A numerical example can be provided by considering the supermassive black hole $M87^*$ with a mass of $\SI{6.5e9}{M_\odot}$ with the same configuration depicted in Fig.~\ref{Schwarzschild} with the source located at $(r_s=30m, \theta_s=\pi/3, \phi_s=\pi/4)$ and observer on the equatorial plane.  In this scenario, after the appearance of the first image, there is a waiting period of 3.5 days for the image on the opposite side, and 12.1 days for the one on the same side.

For comparison, considering the supermassive black hole of our own galaxy Sgr A$^*$, with a mass approximately a thousand times smaller, the time delays are of the order of minutes.

\section{First order geodesics}
The next step consists in taking the geodesics equations, always recovered from the results in appendix, to the first order in $a$.

A convenient way to proceed is to add a perturbation to the position angle, the approach parameter and the time delay to their expressions derived in the previous secion for the Schwarzschild case. We will write
\begin{equation}
    \begin{cases}
        \alpha=\alpha_0+a \ \alpha_1
        \\
        \delta=\delta_0+a \ \delta_1
        \\
        \Delta t_{i,j}=\Delta {t_0}_{i,j}+a \ \Delta {t_1}_{i,j}
    \end{cases}
    \label{ansatz}
\end{equation}
where the terms with the subscript $0$ are the ones we wrote in the preceding section and the subscripts $i$ and $j$ indicate the pair of images between which we calculate the time delay. Similarly, we indicate by $\psi_0$ the scalar deflection derived in Eq.\eqref{psi0}.

Now, substituting the ansatz in the geodesics equations, expanding to the first order in $a$ and then using the zero order results, we obtain
\begin{equation}
\begin{cases}
    \frac{2\delta_1}{\delta_0}-\frac{\alpha_1\sqrt{1-\mu_o^2}\cos{\alpha_0}}{1-(1-\mu_o^2)\cos^2{\alpha_0}}\left[(-1)^n\mu_o\mp\mu_s \tan{\alpha_0}\cot{\alpha_{s 0}}\right]=0
    \\
    \Tilde{\psi}+\frac{\alpha_1}{1-(1-\mu_o^2)\cos^2{\alpha_0}}\left[(-1)^n\mu_o\mp\mu_s \tan{\alpha_0}\cot{\alpha_{s 0}}\right]=0
    \\
    c\Delta {t_1}_{i,j}=6\sqrt{3}m \left(\frac{\delta_{j 1}}{\delta_{j 0}}-\frac{\delta_{i 1}}{\delta_{i 0}}\right)
    \label{firstordergeo}
\end{cases}  
\end{equation}

where
\begin{equation}
   \Tilde{\psi}=\frac{1}{2m}\left(\frac{4}{3\sqrt{3}}\psi_0-\log\frac{S_2(\eta_s)}{S_2(0)}-\log\frac{S_2(\eta_o)}{S_2(0)}\right).
\end{equation}

The linearity of the system above allows to immediately solve it for the perturbation of the position angle and the approach parameter induced by the specific angular momentum with respect to the spherically symmetric case:
\begin{equation}
    \alpha_1=-\Tilde{\psi}\frac{1-(1-\mu_o^2)\cos^2{\alpha_0}}{(-1)^n\mu_o\mp\mu_s \tan{\alpha_0}\cot{\alpha_{s 0}}}
    \label{alfa1}
\end{equation}
\begin{equation}
    \delta_1=-\frac{1}{2}\delta_0\Tilde{\psi}\sqrt{1-\mu_o^2}\cos{\alpha_0}
    \label{delta1}
\end{equation}
As regards the time delay, we again should make the distinction of the case of consecutive images appearing on the same side of the black hole and the case of images appearing on different sides of the black hole with the same number of inversion points in the polar motion.

In the first case, the calculation is straightforward because both $\delta_{j 1}$ and $\delta_{i 1}$ come with the same sign of $\cos{\alpha_0}$. Therefore, one gets:
\begin{equation}
    c\Delta {t_1}_{+}=4\pi\sqrt{1-\mu_o^2}\cos{\alpha_0}
    \label{deltat1}
\end{equation}
In the second case, instead, it is necessary to consider that in $\delta_{i 1}$ (the negative parity image) $\cos{\alpha_0}$ must be taken with different signs with respect to the other one. With this in mind, we obtain:
\begin{equation}
    c\Delta {t_1}_{-}=-4 n \pi \sqrt{1-\mu_o^2}\cos{\alpha_0}
\end{equation}
where $\alpha_0$ now represents only the position angle of the positive parity image.

The key distinction from the previous scenario is that while $\alpha_0$ does not depend on the number of inversion points $n$, $\alpha_1$ does. This dependence is explicitly manifested in the term $(-1)^n$, and also hidden inside $\psi_0$ (see Eq. (\ref{psinpi})). This implies that photons orbiting the black hole undergo the phenomenon of frame dragging, causing each image to move consistently with the rotation. In our illustration, the projection of the spin axis coincide with the axis $r_o\theta_2$, resulting in counterclockwise rotation of the images. Moreover, as the value of $n$ increases, the scalar deflection also increases, making higher-order images more spaced apart. 

As evident from Eqs.\eqref{delta1} and \eqref{deltat1}, these expressions vanish under two specific conditions: when $\mu_o=1$ and when $\alpha_0=\pi/2$. In the first scenario, corresponding to an observer situated along the axis of symmetry, the rotation solely influences the polar angle at which the images manifest, leaving the distance from the shadow border and the time delay unchanged. The second case arises when the source lies in the plane perpendicular to the accretion disk and encompasses the line of sight. In this situation as well, neither the approach parameter nor the time delay is impacted by frame dragging.

In situations where the aforementioned specific conditions are not met, the sign of the perturbations $\delta_1$ and $\Delta t_1$ is determined by $\cos{\alpha_0}$. Co-rotating photons with $\cos{\alpha_0}<0$ exhibit an approach parameter slightly larger than $\delta_0$, while the time delay $\Delta t_+$ is smaller. This behavior is attributed to the frame-dragging effect. Co-rotating photons experience a drag from the spacetime, leading to a reduced effective deflection and consequently a shorter travel time along their trajectory. On the contrary, the behaviour of $\Delta t_-$ is opposite since it deals with both parity images.

\begin{figure}
\centering
\includegraphics[width=0.9\textwidth]{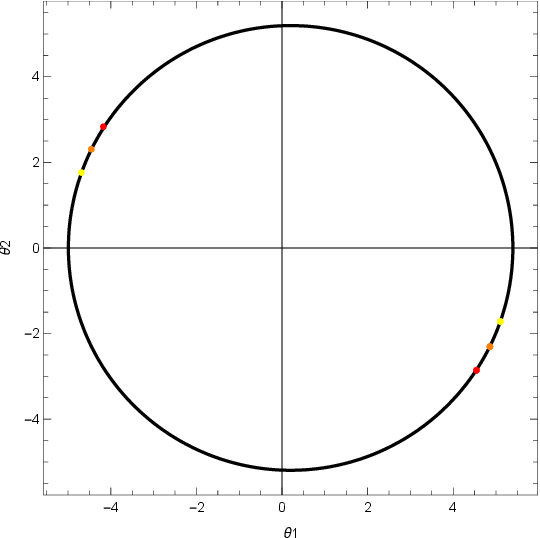}
\caption{The dots represent the relativistic images of order 1,2 and 3 (red, orange and yellow) characterized by the index $n$ being equal to 3, 5 and 7 respectively. The physical situation is the same that we considered in Fig.\ref{Schwarzschild}. The specific angular momentum used to generate this plot is $a=0.1m$ in order to be coherent with the perturbative expansion.}
\label{kerr}
\end{figure}

Finally, both $\delta_1$ and $\Delta t_1$ appear to be of the same order of magnitude with respect to their zero-order values. Consequently, the magnitude of the perturbation is primarily determined by the specific angular momentum $a$. This is the reason why $a=0.1m$ is used in Figs.~\ref{Schwarzschild} and \ref{kerr} as the maximum value for the specific angular momentum to maintain consistency with the perturbative expansion.

To sum up:
\begin{itemize}
    \item collecting the results from Eqs.\eqref{alfa0} and \eqref{alfa1} we have the total position angle
    \begin{equation}
        \alpha=\alpha_0+a\alpha_1;
    \end{equation}

    \item the approach parameter is recovered using Eqs.\eqref{eq:deltazeroquadro} and \eqref{delta1} giving
    \begin{equation}
        \delta^2=\delta_0^2\left(1-\frac{1}{2}a\Tilde{\psi}\sqrt{1-\mu_o^2}\cos{\alpha_0}\right)^2;
    \end{equation}

    \item as before, $\epsilon$ is found inverting Eq.\eqref{sdlparametrization3}. Expanding the result up to the first order in $a$ we have
    \begin{equation}
        \epsilon=\delta_0^2\left[\frac{3}{2}-a\sqrt{1-\mu_o^2}\cos{\alpha_0}\left(\frac{3}{2}\Tilde{\psi}+\frac{1}{\sqrt{3}m}\right)\right];
    \end{equation}
\end{itemize}
With $\alpha$ and $\epsilon$ so determined, we can go back to our parameterization \eqref{sdlparametrization0} of the observer's sky to obtain the position of the images. Fig.~\ref{kerr} provides an example of how the images may appear in this situation.

Concerning the perturbation of the time delay in the Kerr black hole scenario, we have
\begin{equation}
     \Delta t_+=6\sqrt{3}\pi\frac{m}{c}\left(1+\frac{2}{3\sqrt{3}}\frac{a}{m}\sqrt{1-\mu_o^2}\cos{\alpha_0}\right)
\end{equation}
for consecutive images appearing on the same side of the black hole and
\begin{equation}
     \Delta t_-=6\sqrt{3}\pi\frac{m}{c}\left(\frac{h(\alpha_0)}{\pi}-\frac{2}{3\sqrt{3}}\frac{a}{m}n\sqrt{1-\mu_o^2}\cos{\alpha_0}\right)
\end{equation}
for images appearing on opposite sides of the black hole.

The first-order effect of spin on the positions of relativistic images can be further explored in the plots proposed in the following figures.

\begin{figure}
\centering
\includegraphics[width=\textwidth]{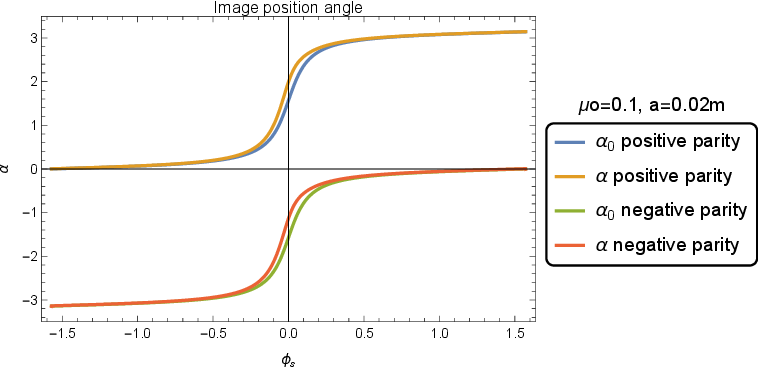}
\caption{Position angle of the $n=3$ images as a function of the source azimuth $\phi_s$ for the Schwarzschild limit ($\alpha_0$) and for a Kerr black hole with spin a =0.02m ($\alpha$). Here we considered the source in the equatorial plane with $\eta_s=0.9$ while the observer is at infinity, with inclination given by $\mu_o = \cos \theta_o = 0.1$.}
\label{fig:alfa}
\end{figure}

Fig.~\ref{fig:alfa} shows the position angle as a function of the source azimut $\phi_s$ for a source on the equatorial plane $\mu_s=0$. When $\phi_s=0$, the source is nearly behind the black hole (the observer is at $\mu_o=0.1$ in this figure, slightly above the equatorial plane). When the source is on the right ($\phi_s<0$) the positive parity image appears on the right ($\alpha=0$) and the negative parity image on the left ($\alpha=-\pi$). Then the two images swing above and below the black hole respectively as the source moves behind ($\phi_s=0$) and finally to the left ($\phi_s>0$). We can notice the perturbation in the position angle induced by the black hole spin, always pushing the image to higher $\alpha$, corresponding to a counterclockwise shift. The shift is maximum at $\phi_s=0$, when the photon trajectory is polar and the source is closer to the zero-order caustic point.

\begin{figure}
\centering
\includegraphics[width=\textwidth]{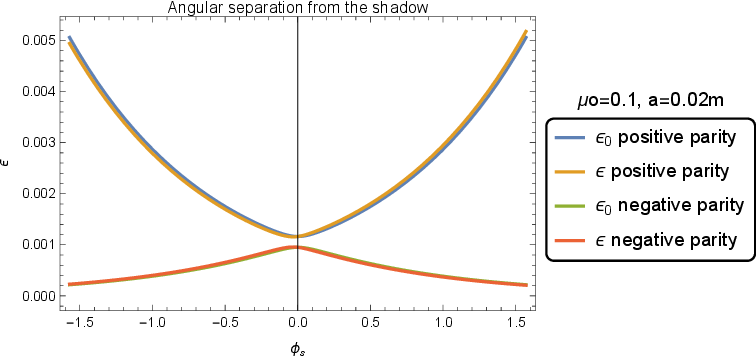}
\caption{Angular separation of $n=3$ images from the shadow border. }
\label{fig:epsilon}
\end{figure}

Fig.~\ref{fig:epsilon} shows the relative angular separation of the images from the shadow boundary as a function of the source azimut $\phi_s$, similarly to the previous figure. For negative values of $\phi_s$, the positive parity image is created by counter-rotating light rays. Consequently, the image appears slightly closer to the shadow border compared to the corresponding image in Schwarzschild spacetime. Conversely, for positive values of $\phi_s$, the positive parity image is formed by co-rotating light rays leading to the opposite effect. Although almost invisible, the effect is completely reversed for negative parity images. Note that $\phi_s=0$ corresponds to a source almost behind the black hole (remember that here $\mu_o = 0.1$), at the closest approach to the caustic point that would lead to the formation of the Einstein ring of order 3.

\begin{figure}
\centering
\includegraphics[width=\textwidth]{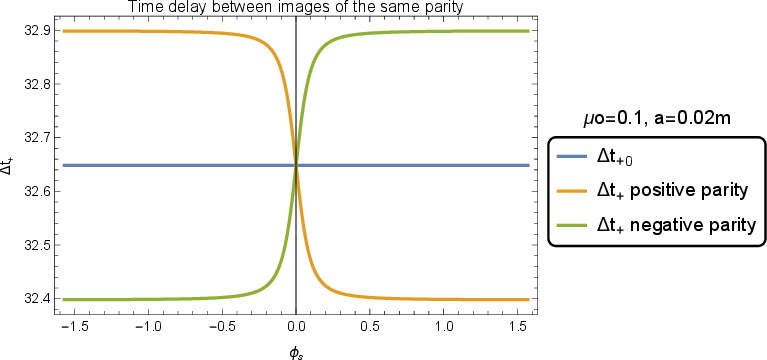}
\caption{Time delay between a couple of images of the same parity. The time delay is measured in unit of $m/c$.}
\label{fig:Deltat+}
\end{figure}

Fig.~\ref{fig:Deltat+} shows the time delay between consecutive images of the same parity. For a non rotating black hole this delay does not depend on $\phi_s$. On the other hand, in Kerr spacetime, when $\phi_s$ is negative the positive parity images are formed by counter-rotating light rays and thus $\Delta t$ increases, when $\phi_s$ is positive the positive parity images are formed by co-rotating light rays causing a decrease in $\Delta t$. The situation is inverted for negative parity images. 

\begin{figure}
\centering
\includegraphics[width=\textwidth]{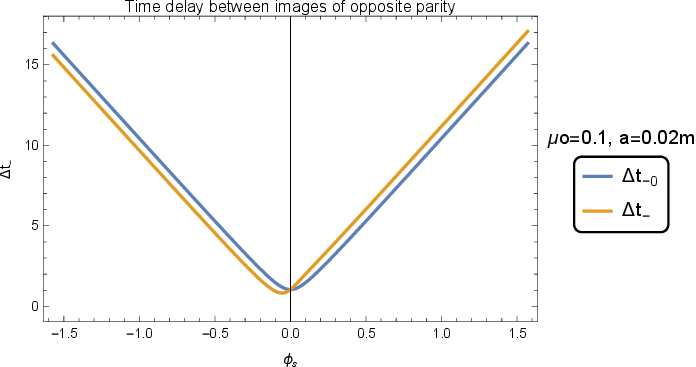}
\caption{Time delay between a negative parity and a positive parity image with the same $n$. Again, the time delay is measured in unit of $m/c$.}
\label{fig:Deltat-}
\end{figure}

Fig.~\ref{fig:Deltat-} shows the time delay between images with opposite parity and the same number of inversion points $n$. The time delay is minimum for $\phi_s=0$ since it would be zero for perfect alignment of source, black hole and observer. $\Delta t_-$ reaches the maximum for $\phi_s=\pm \frac{\pi}{2}$, when negative parity images are formed by photons making a half-tour more than positive parity photons. For negative values of $\phi_s$ the counter-rotating light rays take more time, while the co-rotating light rays take less resulting in a decreased time delay. Opposite situation for positive values of $\phi_s$.

\section{Conclusions}
The EHT provided the possibility of a direct investigation of the regions immediately outside the event horizon using electromagnetic radiation. The accretion disk around supermassive black holes represents an ideal light source in order to study extreme environments and highlight potential effects of the strong gravitational field.

In the next future, with the extension of EHT to space interferometry, the observation of higher order images predicted by the gravitational lensing, the so called photon rings, will become feasible and eventually enlighten our knowledge about how gravity behaves around a black hole.

In this paper we focus on the higher order images of a source whose light suffer gravitational lensing due to a slowly rotating black hole. These images are generated by photons performing one or more loops around the black hole before emerging. Therefore, they carry valuable information about the geometry of space at the so-called photon sphere, which is not far from the black hole horizon. Exploiting the analytical framework of Strong Deflection Limit, which nicely applies to such images, for the first time we develop a first order expansion in the spin parameter that tracks the change in the position of these images that occurs as soon as we turn the black hole angular momentum on. 

It is perhaps unnecessary to underline the advantage of a fully analytic treatment that allows to capture the first order effect of spin with very simple formulae. Such formulae can be extremely useful by themselves to quantify the effects of black hole spin on higher order images in further analytical investigations of Kerr metric and will serve as a reference for other alternative metrics. In addition, numerical ray-tracing codes can take advantage of this analytical limit as a starting point to the description of higher order terms and will serve as a benchmark to test codes in a limit in which analytical formulae perform well. As expected, the limitations of a first order treatment are reached sooner if the source is closer to a caustic point. In this regime, the corrections in the image positions diverges, reminding that a first order treatment ignores the fact that caustics become extended astroids in Kerr lensing \cite{bozza2008}. Here a different second-order treatment becomes necessary \cite{bozza2006kerr}.

Our results show that higher order images as seen by a distant observer rotate counterclockwise around the shadow of the black hole, with an angular shift that scales linearly with the order of the image (as long as the perturbative treatment remains valid). Images become more spaced apart on the co-rotating side and more packed on the counter-rotating side. The time taken by a photon to perform a complete loop is longer for counter-rotating images and shorter for co-rotating images. All these results are coherent with the expectations from basic physics of Kerr black hole and are now immediately visible in the formulae contained in this paper.

\appendix
\section{Resolution of radial integrals}
\label{Resolution of radial integrals}
\setcounter{equation}{0}
\renewcommand{\theequation}{A.\arabic{equation}}
In this appendix we closely follow the SDL technique used in Ref.~\cite{bozza2002gravitational} to solve radial integrals that appear in the geodesics equations, which are
\begin{equation}
     I_1=\int \frac{dr}{\sqrt{R}}
\end{equation}
\begin{equation}
      I_2=\int \frac{r^2+a^2-aJ}{\Delta\sqrt{R}}dr
\end{equation}
\begin{equation}
      I_3=\int \frac{r^2}{\sqrt{R}}dr
\end{equation}
\begin{equation}
      I_4=\int \frac{r(r^2+a^2-a J)}{\Delta\sqrt{R}}d r.
\end{equation}
As already stated, photons follow the path starting from the light source located in $r_s$ to the closest approach distance $r_0$ and then from here to the observer position $r_o$. Remembering that the radial velocity is negative while coming closer to the black hole, both contributions must be summed and in each one of them the integration domain must be $[r_0, r_i]$\footnote{The subscript $i$ is used to unify the observer $o$ and the source $s$ cases.}.

A useful change of variables is the substitution of the radial coordinate $r$ with the new one $\eta$ defined as
\begin{equation}
    \eta=1-\frac{r_0}{r}
    \label{tranformation}
\end{equation}
so that the integration domains will be $[0, \eta_i]$. We keep the notation $\eta_s$ and $\eta_o$ in order to have the most general result but, in any practical case the distance between the lens and the observer is so large with respect to the closest approach distance that $\eta_o$ is always equal to $1$. With this substitution all the integrals acquire the form
\begin{equation}
    I_l=\int f(\eta) g_l(\eta) d\eta
\end{equation}
where 
\begin{equation}
    f(\eta)=\frac{1}{\sqrt{R(\eta)}}
\end{equation}
while the functions $g_l$ include the remaining part of the integrands and the Jacobian of the transformation \eqref{tranformation}. The index $l$, that goes from 1 to 4, is simply used to unify the notation of the four radial integrals above. Writing the functions $g_l$ explicitly we have:
\begin{equation}
    g_1(\eta)=\frac{r_0}{(1-\eta)^2}
\end{equation}
\begin{equation}
    g_2(\eta)=\frac{r_0}{(1-\eta)^2}\frac{r_0^2+a(1-\eta)^2(a-J)}{r_0^2-2m r_0(1-\eta)+a^2(1-\eta)^2}
\end{equation}
\begin{equation}
    g_3(\eta)=\frac{r_0^3}{(1-\eta)^4}
\end{equation}
\begin{equation}
    g_4(\eta)=\frac{r_0^2}{(1-\eta)^3}\frac{r_0^2+a(1-\eta)^2(a-J)}{r_0^2-2m r_0(1-\eta)+a^2(1-\eta)^2}.
\end{equation}

Since the closest approach distance $r_0$ is a zero for the function $R(r)$, this means that $\eta=0$ is a zero for $R(\eta)$ and thus, we can expand this function in a neighborhood of $\eta=0$ up to the second order in $\eta$ obtaining
\begin{equation}
    R(\eta)=\beta_1 \eta +\beta_2\eta^2+o(\eta^2)
\end{equation}
where the coefficients of the expansion are
\begin{equation}
    \beta_1=r_0 \left[2m\left((a-J)^2+Q\right)+2(a^2-J^2-Q)r_0+4r_0^3\right]
\end{equation}
\begin{equation}
    \beta_2=r_0\left[2m\left((a-J)^2+Q\right)+3(a^2-J^2-Q)r_0+10r_0^3\right].
\end{equation}
The expansion itself defines the function
\begin{equation}
    f_0(\eta)=\frac{1}{\sqrt{\beta_1\eta +\beta_2\eta^2}}
\end{equation}
that basically represent the divergence of $f(\eta)$. So, if we sum and subtract inside each radial integral the quantity $f_0(\eta) g_l(0)$, we can split them in a term containing the divergence and in a regular one
\begin{equation}
    I_l=I_{l,D}+I_{l,R}
\end{equation}
where
\begin{equation}
    I_{l,D}=\int f_0(\eta) g_l(0) d\eta
\end{equation}
\begin{equation}
    I_{l,R}=\int [f(\eta)g_l(\eta)-f_0(\eta) g_l(0)] d\eta.
\end{equation}
The significant advantage of this method is that the divergent integral can be evaluated through straightforward analysis, and it yields the following result:
\begin{equation}
    I_{l,D}=\frac{2 g_l(0)}{\sqrt{\beta_2}}\log\frac{\sqrt{\beta_2\eta}+\sqrt{\beta_1+\beta_2\eta}}{\sqrt{\beta_1}}
\end{equation}
where all the quantities appearing in this result are known function of $r_0$, $J$, $Q$ and $a$. Subsequently, we substitute these quantities in favor of the angle $\alpha$ and the approach parameter $\delta$ through the SDL parametrization Eq.\eqref{sdlparametrization1}, Eq.\eqref{sdlparametrization2} and Eq.\eqref{sdlparametrization3}. Eventually, in the limit of vanishing $\delta$, the result is reduced to a term proportional to $\log(\delta)$ plus a constant.

In contrast, the other integrals $I_{l,R}$ is regular in the entire integration domain. Consequently, sending $\delta$ to zero, these integrals do not diverge. This implies that their contribution to the SDL expansion is limited to a constant term and higher-order terms in $\delta$, which can be neglected. 

As regards the calculation, we notice that it is advantageous to perform the leading order expansion in $a$ before the integration. This approach results in a summation of readily integrable functions. Finally, we combine this outcome with the previous one to reconstruct the complete SDL formulas for the radial integrals. All of them appear in the following form:
\begin{equation}
     I_l=-a_l\log\delta +b_{l s} +b_{l o}
\end{equation}
where
\begin{equation}
    a_1 =\frac{2}{3\sqrt{3}}\frac{1}{m}-\frac{4}{27}\frac{a}{m^2}\sqrt{1-\mu_o^2}\cos{\alpha}
\end{equation}
\begin{equation}
    a_2=\frac{2}{\sqrt{3}}\frac{1}{m}-\frac{2}{3}\frac{a}{m^2}\sqrt{1-\mu_o^2}\cos{\alpha}
\end{equation}
\begin{equation}
    a_3 =2\sqrt{3}m+\frac{4}{3}a\sqrt{1-\mu_o^2}\cos{\alpha}
\end{equation}
\begin{equation}
    a_4 =2\sqrt{3}-\frac{2}{3}\frac{a}{m}\sqrt{1-\mu_o^2}\cos{\alpha}
\end{equation}
\begin{equation}
        b_{1i}=\frac{1}{2}a_1\log\left[12 S_{\sqrt{3}} (\eta_i)\right]
\end{equation}
\begin{equation}
    \begin{split}
        b_{2i}=\frac{1}{2}a_2\log\left[12S_{\sqrt{3}} (\eta_i)\right] + \frac{1}{2m}
        \log (1+2\eta_i)S_2(\eta_i)S_2(0)+ \\ 
        -\frac{a}{m^2}\sqrt{1-\mu_o^2}\cos{\alpha}\left[1-\frac{\sqrt{3-2\eta_i}(1+\eta_i)}{\sqrt{3}\left(1+2\eta_i\right)} \right]\qquad\qquad 
    \end{split}
\end{equation}
\begin{equation}
    \begin{split}
        b_{3i}=\frac{1}{2}a_3 \log\left[12S_{\sqrt{3}} (\eta_i)\right]+\left(3m+\frac{2}{\sqrt{3}}a\sqrt{1-\mu_o^2}\cos{\alpha}\right)\left(\frac{\sqrt{3-2\eta_i}}{1-\eta_i}-\sqrt{3}\right)  
    \end{split}
\end{equation}
\begin{equation}
\small
    \begin{split}
        b_{4i}=\frac{1}{2}a_4 \log\left[12S_{\sqrt{3}} (\eta_i)\right]+\log\frac{S_{\frac{5}{3}}(0)}{S_{\frac{5-2\eta_i}{3}}(\eta_i)} 
        -\frac{1}{\sqrt{3}}\frac{a}{m}\sqrt{1-\mu_o^2}\cos{\alpha}\left(\frac{\sqrt{3-2\eta_i}}{1-\eta_i}-\sqrt{3}\right)
    \end{split}
\end{equation}
Here we have also used the function $S_c(\eta)$ defined in Eq. (\ref{Sdefinition}) to shorten the formulae.

\section{Resolution of angular integrals}
\label{Resolution of angular integrals}
\setcounter{equation}{0}
\renewcommand{\theequation}{B.\arabic{equation}}
The angular integrals are
\begin{equation}
     J_1=\int \frac{d\theta}{\sqrt{\Theta}}
\end{equation}
\begin{equation}
        J_2=\int\frac{\csc^2{\theta}}{\sqrt{\Theta}}d\theta 
\end{equation}
\begin{equation}
       J_3=\int\frac{\cos^2{\theta}}{\sqrt{\Theta}}d\theta.
\end{equation}
For all of them, it is convenient to introduce the variable $\mu=\cos{\theta}$ as we already said. With this substitution the integrals become
\begin{equation}
     J_1=\int \frac{d\mu}{\sqrt{\Theta_\mu}}
\end{equation}
\begin{equation}
        J_2=\int\frac{d\mu}{(1-\mu^2)\sqrt{\Theta_\mu}}
\end{equation}
\begin{equation}
       J_3=\int\frac{\mu^2}{\sqrt{\Theta_\mu}}d\mu,
\end{equation}
where
\begin{equation}
   \Theta_\mu=a^2(\mu_-^2+\mu^2)(\mu_+^2-\mu^2) 
\end{equation}
\begin{equation}
    \mu_\pm^2=\frac{\sqrt{b_{JQ}^2+4a^2Q_m}\pm b_JQ}{2a^2}
\end{equation}
\begin{equation}
    b_{JQ}=a^2-J_m^2-Q_m.
\end{equation}
In the previous two relations, we already replaced the constants of motion $J$ and $Q$ with their critical values corresponding to the photon sphere $J_m$ and $Q_m$. This is justified since they contain the quantity $\epsilon$, or equivalently $\delta^2$, but in spirit of the SDL we retain only the  contribution that goes to infinity or at least that are constant in the limit $\delta \to 0^+$.

Looking at the form of the function $\Theta_\mu$, we can clearly notice that it has two zeros in $\mu=\pm \mu_+$. These zeros represent the inversion points in the polar motion: photons perform symmetric oscillations of amplitude $\mu_+$ with respect to the equatorial plane.

In order to better understand how the angular motion works, it is useful to write explicit expression of $\mu_+$ expanded up to the second order in $a$.
\begin{equation}
    \mu_+=\sqrt{1-(1-\mu_o^2)\cos^2{\alpha}}+\frac{1}{54}\frac{a^2}{m^2}\frac{(1-\mu_o^2)^2\cos^2{\alpha}\sin^2{\alpha}}{\sqrt{1-(1-\mu_o^2)\cos^2{\alpha}}}
\end{equation}
In this approximation, the oscillation amplitude is $\sqrt{1-(1-\mu_o^2)\cos^2{\alpha}}$, plus corrections only due to the second order terms in the black hole spin. The minimal amplitude of the oscillations is obtained for the highest value of $\cos{\alpha}$, which gives $\lvert \mu_o\rvert$. This means that purely equatorial trajectories with $\mu_+=0$ are considered in the problem only if the observer itself lies on the equatorial plane or, in other words, only an equatorial observer can detect photons travelling in the equatorial plane.

On the other hand, polar photons that have $\cos{\alpha}=0$ perform oscillations with maximal amplitude $\mu_+=1$. This means that this kind of light rays complete orbits passing above the poles of the black hole.

We can take into account the oscillating nature of the angular motion and simplify the integrals further by introducing the variable $z=\mu/\mu_+$. As a consequence the integrals become:
\begin{equation}
     J_1=\int \frac{d z}{\sqrt{\Theta_z}}
\end{equation}
\begin{equation}
        J_2=\int\frac{d z}{(1-\mu_+^2z^2)\sqrt{\Theta_z}}
\end{equation}
\begin{equation}
       J_3=\int\frac{\mu_+^2z^2}{\sqrt{\Theta_z}}d z,
\end{equation}
where
\begin{equation}
   \Theta_z=a^2(\mu_-^2+\mu_+^2z^2)(1-z^2). 
\end{equation}
In addition, a useful strategy to face these calculations is expanding the integrand functions to the leading order in $a$ before the integration. The primitive functions read
\begin{equation}
\small
\begin{split}
     F_{J_1}(z)=\frac{1}{3\sqrt{3}}\arcsin{z}\left\lbrace \frac{1}{m} -\frac{2}{3\sqrt{3}}\frac{a}{m^2}\sqrt{1-\mu_o^2}\cos{\alpha} \right\rbrace 
\end{split}
\end{equation}
\begin{equation}
\small
    \begin{split}
     F_{J_2}(z)=\frac{1}{3\sqrt{3}\sqrt{1-\mu_o^2}\cos{\alpha}}\arctan\left(\frac{z\sqrt{1-\mu_o^2}\cos{\alpha}}{\sqrt{1-z^2}}\right) \left\lbrace \frac{1}{m} -\frac{2}{3\sqrt{3}}\frac{a}{m^2}\sqrt{1-\mu_o^2}\cos{\alpha}\right\rbrace
\end{split}
\label{fj2}
\end{equation}
We do not report the result of $J_3$ since in the geodesic equations this integral is already multiplied by $a^2$ and thus its contribution is beyond our first order approximation. 

As already mentioned, the integration should cover the whole photon trajectory, which may perform one or more oscillations around the equatorial plane.

In order to make the picture clearer, let us start by considering a photon starting its motion from the source position $z_s=\mu_s/\mu_+$ with growing angular coordinate. 
The first piece of the angular integrals covers the trajectory between the starting point and the upper inversion point: the corresponding domain of integration will be $[z_s,1]$. Subsequently, completing a certain number of orbits around the black hole, the photon will move back and forth between the inversion points. So, indicating with $n$ the number of inversion point touched during its motion, we will have $n-1$ integrals covering the whole domain $[-1,1]$. All these contributions will come with the same sign. Finally, if $n$ is even,
the photon reaches the observer from below so that the last piece of the integral will cover the domain $[-1,z_o]$, otherwise $z$ is decreasing and the domain is $[z_o,1]$. 

The total angular integrals are thus given by the sum of all these contributions and, using the index $l$ assuming the values 1 or 2 to indicate the corresponding integral, we can write 
\begin{equation}
    J_l=F_{J_l}(1)-F_{J_l}(z_s)+(n-1)\left[F_{J_l}(1)-F_{J_l}(-1)\right]+F_{J_l}(z_o)-F_{J_l}(-1)
\end{equation}
for even $n$ and
\begin{equation}
    J_l=F_{J_l}(1)-F_{J_l}(z_s)+(n-1)\left[F_{J_l}(1)-F_{J_l}(-1)\right]+F_{J_l}(1)-F_{J_l}(z_o)
\end{equation}
for odd $n$.

Repeating the same argument but considering the photon leaving the source decreasing its angular coordinate we should write
\begin{equation}
    J_l=F_{J_l}(z_s)-F_{J_l}(-1)+(n-1)\left[F_{J_l}(1)-F_{J_l}(-1)\right]+F_{J_l}(1)-F_{J_l}(z_o)
\end{equation}
for even $n$ and
\begin{equation}
    J_l=F_{J_l}(1)-F_{J_l}(z_s)+(n-1)\left[F_{J_l}(1)-F_{J_l}(-1)\right]+F_{J_l}(z_o)-F_{J_l}(-1)
\end{equation}
for odd $n$.

All the primitive functions $F_{J_l}$ are odd function of $z$ and thus it is possible to express the angular integrals for every case described above, namely for even or odd $n$, in the following compact form:
\begin{equation}
    J_l=\mp\left[F_{J_l}(z_s)-(-1)^n F_{J_l}(z_o)\right]+2nF_{J_l}(1).
\end{equation}

The upper sign refers to photons leaving the source with increasing $z$ and the lower sign to photons emitted with decreasing $z$.

Technically speaking, the primitive function $F_{J_2}$ given by Eq.\eqref{fj2} is not defined in $z=1$. The value $F_{J_2}(1)$ is intended to be calculated in the limit for $z\to 1^-$ which is finite.


\end{document}